\documentclass[prd,twocolumn,aps,nofootinbib,amsmath,amsfonts,amssymb,superscriptaddress,cp]{revtex4-1}
\usepackage[utf8]{inputenc}
\usepackage{graphicx,float}
\usepackage{amsmath,amssymb,amsfonts}
\usepackage{xcolor}
\def\Lag{{\mathcal L}{}}
\def\Sag{{\mathcal S}{}}
\def\Lw{{\stackrel{\bullet}{\mathcal{L}}}{}}
\def\omw{{\stackrel{\bullet}{\omega}}{}}
\def\Tw{{\stackrel{\bullet}{T}}{}}
\def\Kw{{\stackrel{\bullet}{K}}{}}

\begin{document}
	\title{\bf Bulk Action Growth for Holographic Complexity
	}
	\author{ Martin Kr\v{s}\v{s}\'ak }
	\affiliation{Department of Theoretical Physics, Faculty of Mathematics, Physics and Informatics, Comenius University, Bratislava, 842 48, Slovak Republic}
	\email{martin.krssak@gmail.com\\ martin.krssak@fmph.uniba.sk }
	\begin{abstract}
	The action growth proposal relates the holographic complexity to the value of the  action on the Wheeler-de Witt patch.  
	We introduce  a new method of calculating the gravitational action using the ``bulk" term, i.e.  the part of the Einstein-Hilbert action quadratic in connection coefficients.  We demonstrate how to address  the issue of non-covariance of the bulk action and evaluate it using the tetrad formalism.  Due to the boundary term-free nature of the bulk action, we can gain further insights into the spatial structure of the action on the Wheeler-de Witt patch.  We then argue  that our entire scheme can be naturally covariantized within the framework of  teleparallel geometry.
	\end{abstract}
	\maketitle

\section{Introduction}
The AdS/CFT correspondence  established a relation between a black hole in asymptotically AdS spacetime and the dual field theory at its boundary \cite{Maldacena:1997re,Witten:1998qj}.  In addition to studying the properties of black holes outside the horizon, there is currently interest in understanding how the physics is encoded holographically behind the horizons. One  approach is the holographic complexity conjecture, which suggests that the quantum information complexity is given by the value of the gravitational action in the interior region of a black hole
\cite{Brown:2015lvg}\cite{Brown:2015bva}.

The gravitational action in the holographic complexity conjecture is taken to be the standard Einstein-Hilbert action with the cosmological term and the Gibbons-Hawking-York boundary term  \cite{Brown:2015lvg}\cite{Brown:2015bva}
\begin{equation}\label{ghyapp}
	\Sag_\text{grav}=
	\frac{1}{2\kappa} \int_{\mathcal{W}} \sqrt{-g} (R-2\Lambda)
	+
	\frac{1}{\kappa} \int_ {\partial \mathcal{W}} \sqrt{-\gamma} \mathcal{K},
\end{equation}
where $\kappa=8\pi$ (in $c=G=1$ units), $\gamma_{\mu\nu}$ is the  boundary metric, and $\mathcal{K}$ is its extrinsic curvature.

This action is  evaluated in the interior region  of a black hole, $\mathcal{W}$, known as the Wheeler-de Witt (WdW) patch, where the crucial contribution comes from the boundary term. In the case of a  black hole with a single horizon, the WdW patch at late times represents the whole under the horizon region and both  the horizon and $r=0$ are being effectively treated as boundaries. In the case of a black hole with two horizons, the WdW patch at late times corresponds to the region between them and both horizons are treated as boundaries.

We introduce here a new  method of evaluating the gravitational action for the   holographic complexity  by integrating the bulk gravitational action. Our use of the term ``bulk" comes from  the decomposition of the  Einstein-Hilbert Lagrangian  \cite{Landau:1982dva,dInverno:1992gxs,Padmanabhan:2010zzb}
\begin{equation}\label{decomp}
	\Lag_\text{EH}=\Lag_\text{bulk}+\Lag_\text{tot},
\end{equation}
where $\Lag_\text{tot}$ is the total derivative term and the bulk term, $\Lag_\text{bulk}$, is the term  quadratic in connection coefficients\footnote{Note that the term "bulk" is commonly used in the literature to refer to the entire Einstein-Hilbert action, but we refrain from using this term in that way.}.

We motivate our approach by an observation that the total derivative term is not dynamical since it can be transformed into a surface term  using the Stoke's theorem.
The inclusion of the total derivative term introduces second derivatives of the dynamical gravitational field variables, which leads to a problem with the variational principle. The standard solution to this problem is to add the GHY boundary term to cancel out the variations of the total derivative term \cite{York:1972sj,Gibbons:1976ue}, resulting in the total gravitational action \eqref{ghyapp}.

However, there is another solution to avoid the issue of second derivatives by considering the bulk term only. Indeed, it is well-known that both the EH and bulk actions lead to the same field equations and hence are dynamically equivalent \cite{Landau:1982dva, dInverno:1992gxs, Padmanabhan:2010zzb}. 
Our goal in this paper is to demonstrate that this goes beyond the dynamical equivalence and that we can evaluate the bulk action to obtain the same value of the action on the WdW patch as using \eqref{ghyapp} to study holographic complexity of black holes. We show that the boundary term-free nature of the bulk action allows us to study not only the action growth but also the Lagrangian growth, providing us with further insights into the structure of the action on the WdW patch.

While the total derivative term in \eqref{decomp} is not dynamical, it ensures the covariance of the Einstein-Hilbert action. Therefore, if we consider the bulk term alone, we have to face the issue of  non-covariance  of the bulk action. In the standard metric formalism, the bulk term is given by the Einstein Lagrangian \cite{Einstein:1916cd} and  requires evaluation in a peculiar coordinate system, which we argue exists only for the exterior solution and not for the interior solution on the WdW patch that we are interested in. This motivates us then to consider  a tetrad version of the bulk term,  where the issue of non-covariance is not related to the choice of a coordinate system  but to the local Lorentz degrees of freedom of the tetrad. We  show that we can find a well-behaved tetrad on the WdW patch that  gives us the correct answer for the action growth.

We will demonstrate that the issue of the non-covariance can be resolved by reformulating general relativity using the so-called teleparallel geometry. We argue that teleparallel geometry is a natural framework for this task due to the fact that its connection is determined solely by the choice of local Lorentz degrees of freedom, and covariantizes the action without introducing second derivatives of the dynamical field variables.

\section{Bulk actions for general relativity}
The fundamental variable in the metric formalism is the metric tensor $g_{\mu\nu}$, from which we define the Christoffel symbols, $\Gamma^\rho{}_{\nu\mu}$, and the Riemannian curvature tensor leading to the scalar curvature and the Einstein-Hilbert action. The decomposition \eqref{decomp} in the metric case leads us to the metric bulk term \cite{Landau:1982dva,dInverno:1992gxs,Padmanabhan:2010zzb}
\begin{equation}\label{LagEin}
	\Lag_\text{bulk}= \frac{1}{2\kappa}\sqrt{-g}g^{\mu\nu}(
	\Gamma^\rho{}_{\sigma\mu}\Gamma^\sigma{}_{\rho\nu}-
	\Gamma^\rho{}_{\mu\nu}\Gamma^\sigma{}_{\rho\sigma}
	),
\end{equation}
which is commonly known as the Einstein Lagrangian   and defines the Einstein energy-momentum pseudotensor
\cite{Einstein:1916cd,Einstein:1916vd}. 

The Einstein Lagrangian does not transform as a proper 
scalar density under diffeomorphisms and hence its value depends on the choice of a coordinate system. For instance, in  Minkowski spacetime, where gravity is absent and we expect the action to be zero, the Einstein action \eqref{LagEin} indeed  vanishes in  Cartesian coordinates, but  diverges in the spherical coordinate system. 

When including gravity and dealing with a curved spacetime, we must continue using  a quasi-Cartesian coordinate system to avoid the same type of divergent actions. This is closely related to the well-known problem of various pseudotensorial definitions of energy, which are required to be evaluated in such special coordinates \cite{Landau:1982dva}. For example, in the case of Schwarzschild, we  use the isotropic Cartesian coordinate system to find the energy. However, such coordinates are only well-defined in the exterior region of a black hole, and they do not exist in the interior region that we are interested in.

This problem can be easily avoided by using the tetrad formalism and utilizing its coordinate-free nature. We select a set of four orthonormal vectors, $h^a{}_\mu$, called the tetrad  as the fundamental variable,  which is related to the metric by
\begin{equation}\label{met}
	g_{\mu\nu}=\eta_{ab}h^a_{\ \mu} h^b_{\ \nu},
\end{equation}
where $\eta_{ab}=\text{diag}(-1,1,1,1)$ is the tangent space metric.  The covariant differentiation  is done using the Levi-Civita spin connection
\begin{equation}\label{lccon}
	\omega^a{}_{b\mu}=\frac{1}{2} h^{\;c}{}_{\mu} \Big[f_b{}^a{}_c + f_c{}^a{}_b - f^a{}_{bc}\Big],
\end{equation}
where $f^c{}_{a b} = h_a{}^{\mu} h_b{}^{\nu} (\partial_\nu
h^c{}_{\mu} - \partial_\mu h^c{}_{\nu} )$
are the coefficients of anholonomy. The curvature tensor of the spin connection is
\begin{eqnarray}
	R^{a}_{\,\,\, b\mu\nu} = \partial_\mu\omega^{a}_{\,\,\,b\nu}-
	\partial_\nu\omega^{a}_{\,\,\,b\mu}
	+\omega^{a}_{\,\,\,c\mu}\omega^{c}_{\,\,\,b\nu}-\omega^{a}_{\,\,\,c\nu}
	\omega^{c}_{\,\,\,b\mu} \,,
	\label{curvlc}
\end{eqnarray}
from where we define the scalar curvature and the tetrad Einstein-Hilbert action. The decomposition \eqref{decomp} provides us with the tetrad bulk Lagrangian
\begin{equation}\label{BulkLag}
	\Lag_\text{bulk}=	 \frac{1}{2\kappa}h 
	\left(\omega^{a}_{\,\,\,ca}\omega^{bc}{}_{b}
	-\omega^{a}{}_{cb}
	\omega^{bc}{}_{a} 
	\right),
\end{equation}
which was introduced originally by M\o ller and used to define the M\o ller energy-momentum complex \cite{Moller1961,Pellegrini1963,Moller1966,Moller1978}.

To illustrate the problem of non-covariance, we can consider Minkowski spacetime again. The diagonal Cartesian tetrad gives us the expected zero bulk action but the diagonal tetrad in the spherical coordinate system
$h^a{}_\mu=\text{diag}(1,1,r,r\sin\theta)$ leads to a divergent action. 
We can solve this problem by considering a local Lorentz transformation of the tetrad
\begin{equation}\label{proptet}
	\tilde{h}^a_{\ \mu}=\tilde{\Lambda}^a_{\ b} h^b_{\ \mu},
\end{equation}
with a local Lorentz matrix
\begin{equation}\label{specLorentz}
	\tilde{\Lambda}^a_{\ b}=
	\left( \begin{array}{cccc}
		1 & 0 & 0 & 0\\ 
		0 & \cos\phi\sin\theta & \cos\phi\cos\theta & -\sin\phi \\
		0 & -\cos\theta & \sin\theta & 0 \\
		0 & \sin\phi\sin\theta & \sin\phi\cos\theta & \cos\phi
	\end{array}
	\right),
\end{equation}
which will lead to the finite zero action. 

Here, we can observe the usefulness of the tetrad formalism, as it allows us to work both in spherical coordinates and keep the desired properties of quasi-Cartesian coordinates at the same time. The form of the local Lorentz transformation \eqref{specLorentz} can be understood easily from the fact that the  tetrad \eqref{proptet} is obtained from the diagonal  Cartesian tetrad by just a coordinate change  \cite{Krssak:2015lba}.

\section{Bulk Action Growth of AdS Black Holes \label{secgrowth}}
For the holographic complexity conjecture we are interested in evaluating the action\footnote{The integral here is $\int_\mathcal{W} =\int dt\int_0^\pi d\theta \int_0^{2\pi}d\phi\int_{r_-}^{r_+} dr $, where $r_+$ and $r_-$ are boundaries of the WdW patch.}  
\begin{equation}
\Sag=\int_\mathcal{W}\Lag,	
\end{equation}
with the total Lagrangian  given by
\begin{equation}\label{lagtot}
\Lag=\Lag_\text{bulk}+\Lag_\Lambda+\Lag_\text{EM},
\end{equation}
where $ \Lag_\Lambda=\kappa^{-1} h \Lambda=-3\kappa^{-1} h l^{-2}$ is the cosmological term and $\Lag_\text{EM}$ is the Maxwell term.

The spherically symmetric spacetime is given by the metric
\begin{equation}\label{metdiag}
	ds^2 =
	-f^2dt^2 +f^{-2}dr^2 + r^2d\theta^2+ r^2\sin^2 \theta d\phi^2
\end{equation}
that can be represented  by a diagonal tetrad
\begin{equation}\label{tetdiag}
h^a{}_\mu=\text{diag}(f,f^{-1},r,r\sin\theta).
\end{equation}
We encounter a similar issue with this diagonal tetrad, resulting in the same kind of divergence  as discussed in the Minkowski case. Fortunately, it has  the exact same solution even in the curved case, i.e. to use a tetrad \eqref{proptet} with the same  local Lorentz transformation matrix \eqref{specLorentz} but with $h^a{}_\mu$ given by \eqref{tetdiag}.

\subsection{Uncharged case \label{secunchar}}
In the case of the uncharged AdS black hole, we consider the action without the Maxwell term and the $f$-function  given by
\begin{equation}
	f^2=1-\frac{2M}{r}+\frac{r^2}{l^2}. \label{adsmet}
\end{equation}
We find the total Lagrangian \eqref{lagtot} to be 
\begin{equation}\label{lagres}
		\Lag=\frac{2 \sin\theta}{\kappa l^2 r f} [Ml^2+r^3+r(l^2+3r^2)(f-1)].
\end{equation}
By integrating, we obtain the action
\begin{equation}
		\Sag= \int_\mathcal{W} \Lag=\int dt
	\left.	\frac{ r^3+rl^2(1-f)}{l^2} \right|_0^{r_h},
\end{equation}
leading to the action growth
\begin{equation}
\frac{d\Sag}{dt}=2M,	
\end{equation}
which exactly coincides with the result obtained in \cite{Brown:2015bva,Brown:2015lvg} using the standard action \eqref{ghyapp}.

\subsection{Charged case \label{secchar}}
The Reissner-Nordstrom solution is given  by the same metric \eqref{metdiag} and the tetrad \eqref{proptet} as in the uncharged case, but with the  $f$-function given by   
\begin{equation}
f^2=1-\frac{2M}{r}+\frac{Q^2}{r^2} +\frac{r^2}{l^2},\label{rnmet}
\end{equation}
from where two horizons $r_\pm$ are found as the real solutions to $f(r_\pm)=0$

The total Lagrangian, including the Maxwell term for a point charge $\Lag_\text{EM}=\kappa^{-1}h\, Q^2 r^{-4}$, after evaluation takes the same form as in the uncharged case \eqref{lagres} but with the $f$-function given by  \eqref{rnmet}. By integrating, we obtain the bulk action
\begin{eqnarray}
	\Sag&=& \int_{\mathcal{W}} \Lag=\int dt
	\left.	
	\frac{r^3+rl^2(1-f)}{l^2}
	 \right|_{r_-}^{r_+}.
\end{eqnarray}
Using $f(r_\pm)=0$, we find the action growth to be
\begin{equation}
	\frac{d\Sag}{dt}=Q^2 \left(\frac{1}{r_-}-\frac{1}{r_+}\right),
\end{equation}
which is  again  exactly the result obtained in \cite{Brown:2015bva,Brown:2015lvg} using the standard action \eqref{ghyapp}.
\section{Covariantization and teleparallel geometry \label{sectele}}
We would like to demonstrate that the  
teleparallel geometry is a natural framework where  the bulk action can be covariantized. 
The basic idea underlying the teleparallel approach to gravity is to replace the Riemannian torsion-free connection \eqref{lccon} by the curvature-free teleparallel spin connection \cite{Aldrovandi:2013wha,Krssak:2018ywd}\footnote{We use a common convention that geometric objects with respect to the teleparallel connection have $``\bullet"$ above them \cite{Aldrovandi:2013wha}, to distinguish them from the Riemannian geometry quantities. }
\begin{equation}
	\omw^a_{\ b\mu}=\Lambda^a_{\ c} \partial_\mu (\Lambda^{-1})^c_{\ b},
	\label{telcon}
\end{equation}
where $\Lambda^a_{\ c}$ is a local Lorentz transformation. 

The torsion tensor
$\Tw^a_{\ \mu\nu}=
2\partial_{[\mu} h^a_{\ \nu]}+2\omw^a_{\ b[\mu}h^b_{\ \nu]}
$  is  generally non-zero  
and  can be used to formulate a gravity theory by considering the torsion-based Lagrangian \cite{Aldrovandi:2013wha}\footnote{We follow  \cite{Brown:2015lvg} and use $(-,+,+,+)$ convention, unlike \cite{Aldrovandi:2013wha}.}
\begin{eqnarray}
\Lw_\text{TG}=
- \frac{h}{2 \kappa} \left[\Kw^{abc}\Kw_{cba}-\Kw^{ac}{}_{a}\Kw^b{}_{cb}\right], \label{lagtegr}
\end{eqnarray}
where $
\Kw^{a}_{\  bc}=\frac{1}{2}
\left(
\Tw^{\ a}_{c \ b}
+\Tw^{\ a}_{b \ c}
-\Tw^{a}_{\  b c}
\right)$
is the contortion tensor.
Using the Ricci theorem
\begin{equation}\label{Ricci}
	\omw^a_{\ b\mu}=\omega^a_{\ b\mu} + \Kw^a_{\ b\mu},
\end{equation}
it is possible to show that \eqref{lagtegr} is equivalent to the EH action up to a surface term \cite{Aldrovandi:2013wha}
\begin{equation}
	\Lag_{EH}= \Lw_\text{TG} +\partial_\mu \left(\frac{h}{\kappa} \, \Tw^{\nu\mu}{}_\nu \right), \label{lagequiv}
\end{equation}
from where follows that the theory given by \eqref{lagtegr} is dynamically equivalent to general relativity \cite{Aldrovandi:2013wha,Krssak:2018ywd}. 

Moreover, we can relate the teleparallel \eqref{lagtegr} and bulk  \eqref{BulkLag} Lagrangians by using our previous result   \cite{Krssak:2015lba},  where we have derived a relation between $\Lw_\text{TG}(h^a_{\ \mu},\omw^a_{\ b\mu})$ and $\Lw_\text{TG}(h^a_{\ \mu},0)$. Using the Ricci theorem \eqref{Ricci}, it is straightforward to see that  the latter is equivalent to the bulk Lagrangian \eqref{BulkLag}. This leads to the relation
\begin{equation}\label{equivLag}
	\Lw_\text{TG}=\Lag_\text{bulk}+ \frac{1}{\kappa} \partial_\mu \left(
	h \omw^{\nu\mu}{}_\nu
	\right).
\end{equation}

The teleparallel Lagrangian \eqref{lagtegr} is invariant under simultaneous local Lorentz transformation of the tetrad and  spin connection
\begin{equation}
	h^{a}_{\ \mu}\rightarrow\Lambda^a{}_{b}h^{b}_{\ \mu},
\quad
	\omw{}^a_{\ b\mu}\rightarrow \Lambda^a{}_{c} \omw{}^c_{\ d\mu}\Lambda_b^{\ d}+\Lambda^a_{\ c} \partial_\mu \Lambda_b{}^{c},
	\label{spintransf}
\end{equation}
where $\Lambda_b{}^c=(\Lambda^{-1})^c_{\ b}$, and can be viewed as a covariantization of the bulk Lagrangian \eqref{BulkLag}.

Note that while the total derivative term in \eqref{decomp} covariantizes the bulk action by introducing the second derivatives of the dynamical field variables, the total derivative term in \eqref{equivLag} contains only second derivatives of the local Lorentz degrees of freedom  $\Lambda^a{}_b$, which are not dynamical.

\newpage
\onecolumngrid\,
\begin{figure}[t!]
	\begin{center}
		\includegraphics[height=3.5cm]{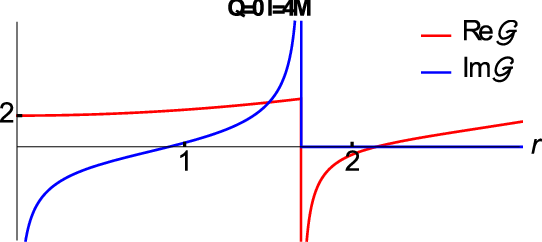}
		\hspace{0.3cm}
		\includegraphics[height=3.5cm]{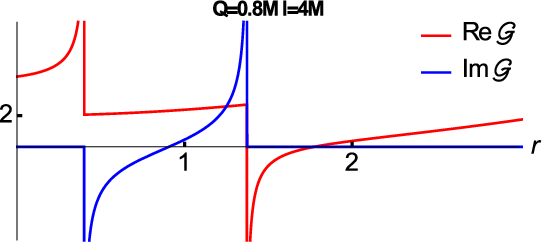}
		\\ \, \\
		\includegraphics[height=3.5cm]{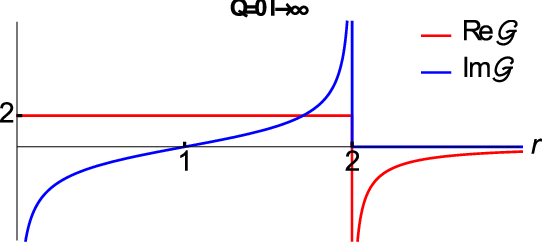}
		\hspace{0.3cm}
		\includegraphics[height=3.5cm]{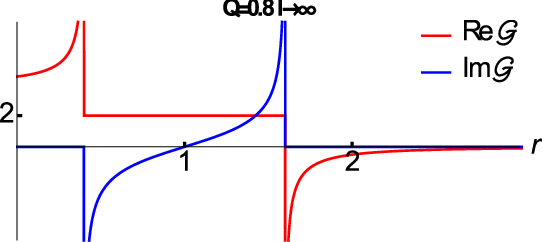}
	\end{center}
	\caption{ The real (red) and imaginary (blue) parts  of the Lagrangian growth $\mathcal{G}=\kappa\frac{d\Lag}{dt}$ ( for $\theta=\pi/2$), where $\Lag$ is given by \eqref{lagres}, for: AdS (upper left), charged AdS  (upper right), Schwarzschild (lower left), Reissner-Nordstr\"om (lower right) black holes.\label{fig1}}
\end{figure}\\
\twocolumngrid
\section{Discussion and Conclusions}

We have introduced a new method of evaluating the gravitational action on the WdW patch at late times, which instead of the standard action \eqref{ghyapp} uses the tetrad bulk action \eqref{BulkLag}, and tested it  on the  asymptotically AdS black holes. In addition to being a useful new tool for calculating the action, our results are interesting from several  perspectives.

The boundary term-free nature of the bulk action allows us to not only calculate the action's value but study the Lagrangian as well. This helps us to gain a deeper understanding of the action's structure on the WdW patch. We plot the ``Lagrangian growth" as a function of distance $r$ in Figure~\ref{fig1}. The upper two plots corresponds to the results for AdS black holes from Section~\ref{secgrowth}. 

We observe a peculiar feature that the Lagrangian becomes complex on the WdW patch. Although the total action is real since the imaginary part is antisymmetric around the center of the WdW patch,  the presence of a complex phase is intriguing and may play a role, for example, in situations where we consider a gravitational path integral \cite{Hawking:1978jz} on a portion of the WdW patch.

It is also interesting to plot the Lagrangian growth for the asymptoticaly flat black holes, which can be seen in the lower two plots in Figure 1. We observe that on the WdW patch, the real part of the Lagrangian is constant. As a result, the growth of the bulk action is determined just by the radial length of the WdW patch. 

Our results are interesting not only in relation to the holographic complexity proposal but also from the perspective of general relativity alone. While the dynamical equivalence of the EH and bulk actions is well-established, the bulk action is considered to be a rather obscure object due to its non-covariance. Similarly,  various pseudotensorial definitions of energy-momentum derived from the bulk action, despite being known to give correct results \cite{Chang:1998wj}, have   fallen out of favor recently.

We have demonstrated that this is mostly unjustified and the bulk action can provide meaningful results as long as we learn how to deal with its non-covariance properly. We have demonstrated this in the tetrad case \eqref{BulkLag}, where we can use spherical coordinates and eliminate divergences through local Lorentz transformation \eqref{specLorentz}. It remains an open question whether we can use the metric bulk Lagrangian \eqref{LagEin} in the same fashion. The isotropic Cartesian coordinates were used successfully in the exterior region
\cite{BeltranJimenez:2018vdo}, but it is not clear how to proceed on the WdW patch.

We have then explored an alternative framework where instead of finding a well-behaved tetrad by applying \eqref{specLorentz} on \eqref{tetdiag}, we  used \eqref{specLorentz} to define a teleparallel spin connection \eqref{telcon}. The torsion tensor of this connection  behaves covariantly under transformations \eqref{spintransf} and  defines a covariant teleparallel gravity \cite{Krssak:2015rqa,Krssak:2018ywd,Krssak:2015oua}. Since the Lagrangian of teleparallel gravity is equivalent to the bulk Lagrangian \eqref{equivLag}, and it is covariant, teleparallel gravity can be viewed as a natural covariant framework for the use of the bulk action.  

From another perspective, we can view our results as that the physics is  given solely by the bulk action. In the standard approach, the bulk term is  covariantized by addition of the total derivative term \eqref{decomp}, but this introduces second derivatives of dynamical field variables that need to be removed by the GHY term. Teleparallel gravity  achieves covariance of the bulk action  by  using  the teleparallel connection \eqref{telcon}, which only contains non-dynamical local Lorentz degrees of freedom that fully remove divergences of the bulk action caused by its non-covariance.

Furthermore, this  is useful also from a rather practical perspective as it allows us to utilize many results already existing in the literature on teleparallel gravity in a new way. In fact, we have here just used our previous results  about the teleparallel action in the exterior region of a black hole \cite{Krssak:2015rqa,Krssak:2015lba}  and applied them to the interior geometry. In recent years, there has been  significant interest in various modified teleparallel theories of gravity, such as $f(T)$ gravity \cite{Ferraro:2006jd,Ferraro:2008ey}, where one of the central problems is finding  well-behaved tetrads  \cite{Ferraro:2011ks,Tamanini:2012hg,Hohmann:2019nat,Coley:2019zld,Emtsova:2021ehh} (See reviews \cite{Cai:2015emx,Krssak:2018ywd,Bahamonde:2021gfp} as well). While these modified theories are rather speculative in their nature, the  results regarding these well-behaved tetrads are  valid on their own and can now   be utilized towards studying holographic complexity using the bulk action growth.

Our proposal was tested here only in the limit of late boundary times, where the WdW patch corresponds to the region according to our definition. However,  action growth was studied in more general situations, including more complicated null boundaries \cite{Lehner:2016vdi} and the WdW patch at intermediate times \cite{Carmi:2016wjl,Carmi:2017jqz}. It would be interesting to apply  our bulk action method  in these more general scenarios, which could turn out to be non-trivial tests of our proposal. However, the main difficulty is that as we move away from the limit of WdW patch at late times, we need to integrate over discontinuous and singular functions, as seen in Fig.\ref{fig1}, which cannot be done straightforwardly and requires further investigation.\\

\section{Acknowledgements}
This work was funded through  SASPRO2 project \textit{AGE of Gravity: Alternative Geometries of Gravity}, which has received funding from the European Union's Horizon 2020 research and innovation programme under the Marie Skłodowska-Curie grant agreement No. 945478.

\end{document}